\documentclass[conference]{IEEEtran}
\IEEEoverridecommandlockouts
\usepackage[ruled,linesnumbered]{algorithm2e}
\usepackage{amsmath,amsfonts}
\usepackage{cases}
\usepackage{mathtools}
\usepackage{amssymb}
\usepackage{bm}
\usepackage{relsize}
\usepackage{array}
\usepackage{textcomp}
\usepackage{float}
\usepackage{stfloats}
\usepackage{url}
\usepackage{verbatim}
\usepackage{graphicx}
\usepackage{subfigure}
\usepackage{cite}
\usepackage{color}
\usepackage{makecell}
\usepackage{etoolbox}
\usepackage{xparse}
\newtheorem{proposition}{\underline{Proposition}}

\newtheorem{lemma}{\underline{Lemma}}

\allowdisplaybreaks[4]

\usepackage{microtype}
\begin{document}

	\title{Optimal Transmit Beamforming for MIMO ISAC with Unknown Target and User Locations}

\author{
	\IEEEauthorblockN{Yizhuo Wang and Shuowen Zhang}
	\IEEEauthorblockA{	Department of Electrical and Electronic Engineering, The Hong Kong Polytechnic University\\[-2pt]
	E-mail: yizhuo-eee.wang@connect.polyu.hk, shuowen.zhang@polyu.edu.hk}
	\thanks{This work was supported in part by the National Natural Science Foundation of China under Grant 62471421, in part by the General Research Fund from the Hong Kong Research Grants Council under Grant 15230022, and in part by the Young Collaborative Research Grant from the Hong Kong Research Grants Council under Grant PolyU C5002-23Y.}
}

\maketitle
\vspace*{-30pt}

\begin{abstract}
This paper studies a challenging scenario in a multiple-input multiple-output (MIMO) integrated sensing and communication (ISAC) system where the locations of the sensing target and the communication user are both \emph{unknown} and \emph{random}, while only their probability distribution information is known. In this case, how to fully utilize the spatial resources by designing the transmit beamforming such that both sensing and communication can achieve satisfactory performance statistically is a difficult problem, which motivates the study in this paper. Moreover, we aim to reveal if it is desirable to have similar probability distributions for the target and user locations in terms of the ISAC performance. Firstly, based on only probability distribution information, we establish communication and sensing performance metrics via deriving the expected rate or \emph{posterior Cram\'{e}r-Rao bound (PCRB)}. Then, we formulate the transmit beamforming optimization problem to minimize the PCRB subject to the expected rate constraint, for which the optimal solution is derived. It is unveiled that the rank of the optimal transmit covariance matrix is upper bounded by the summation of MIMO communication channel matrices for all possible user locations. Furthermore, due to the need to cater to multiple target/user locations, we investigate whether dynamically employing different beamforming designs over different time slots improves the performance. It is proven that using a static beamforming strategy is sufficient for achieving the optimal performance. Numerical results validate our analysis, show that ISAC performance improves as the target/user location distributions become similar, and provide useful insights on the BS-user/-target association strategy. \looseness=-1
\end{abstract}

\vspace{-2mm}
\section{Introduction}\label{sec_int}
\vspace{-1mm}

The evolution towards sixth-generation (6G) networks is marked by a shift from traditional communication to multi-functional platforms enabled by integrated sensing and communication (ISAC) \cite{saad2020vision}. This technology is critical for supporting critical applications such as industrial internet-of-things (IIoT) and low-altitude economy. \looseness=-1

To harness the fully potential of ISAC, judicious design of the transmit signal is of paramount importance. Existing works have primarily considered two sensing performance metrics. The first is the similarity between the achieved and desired beampatterns \cite{hua2023optimal}, which is tractable but does not directly capture sensing errors. The second is the Cramér-Rao bound (CRB), which provides a lower bound on the sensing mean-squared error (MSE) and has been widely studied \cite{liu2022cramer,hua2024mimo}. However, CRB depends on the \emph{exact values} of the parameters to be sensed, which are generally \emph{unknown a prior} in practical sensing scenarios. \looseness=-1

In practice, the parameters to be sensed are often \emph{unknown} and \emph{random}, while their statistical distributions can be known \emph{a priori} from target properties and historical data \cite{xu2024mimo,xu2024isac,yao2025optimal,hou2023optimal,attiah2023active,liu2024ris,ghaddar2025active}. Leveraging this prior information, the \emph{posterior Cram\'{e}r-Rao bound (PCRB)} or Bayesian Cram\'{e}r-Rao bound (BCRB)
characterizes a lower bound of the MSE that depends only on the probability density function (PDF) \cite{van1968detection}. Recent works have used PCRB as the sensing performance metric to optimize transmit signals in multi-antenna sensing and ISAC systems \cite{xu2024mimo,xu2024isac,yao2025optimal,hou2023optimal,attiah2023active,liu2024ris,ghaddar2025active}. For example, \cite{xu2024mimo} characterized the PCRB for multiple-input multiple-output (MIMO) ISAC and derived the optimal transmit covariance matrix, where a novel \emph{probability-dependent power focusing} effect was also revealed. Furthermore, \cite{xu2024isac,yao2025optimal,hou2023optimal} studied the beamforming optimization for single-/multi-user multiple-input single-output (MISO) ISAC, respectively, where the numbers of dedicated sensing beams needed were also unveiled. \looseness=-1

\begin{figure}[t]
	\centering
	\vspace{-8mm}
	\includegraphics[width=0.40\textwidth]{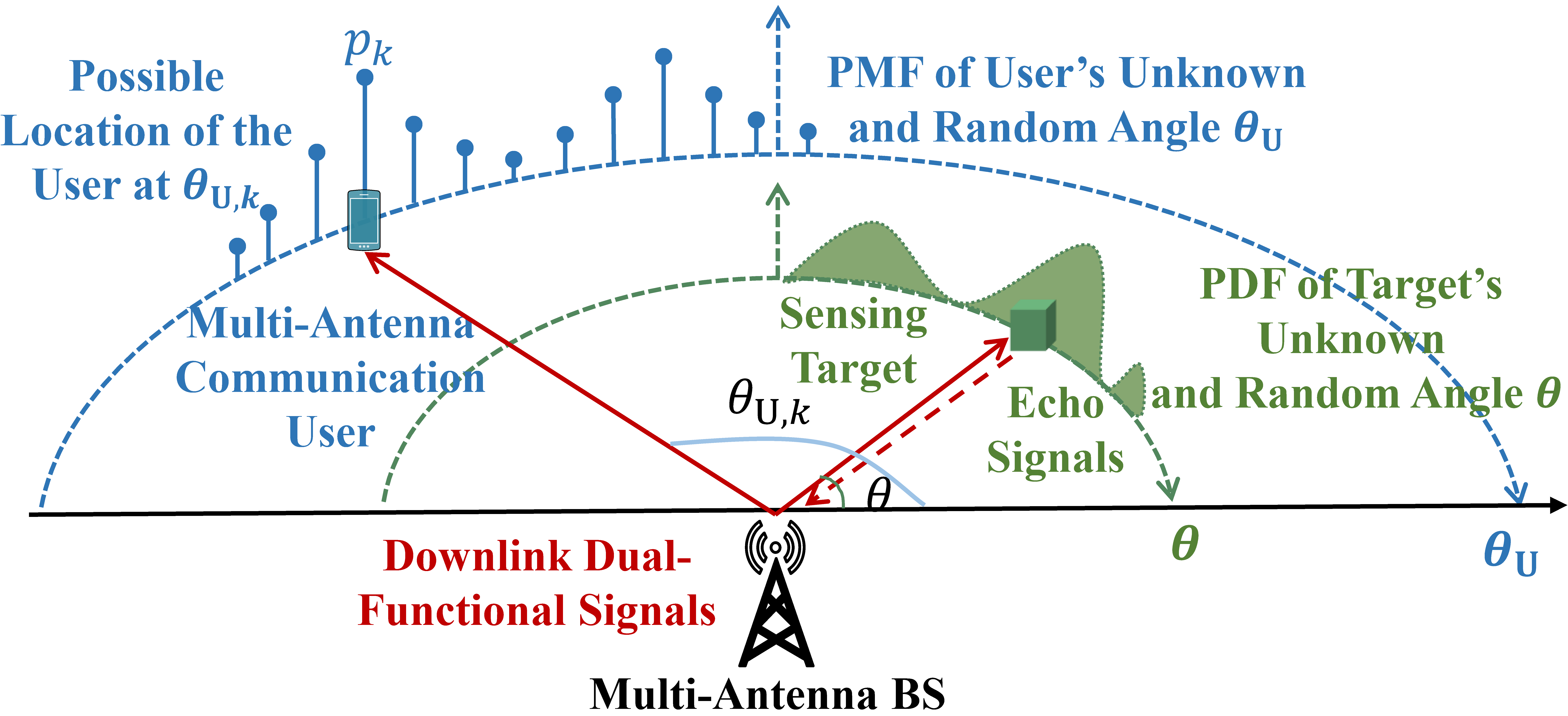}
	\vspace{-4mm}
	\caption{A MIMO ISAC system with unknown target and user locations.}
	\label{system model}
\end{figure} 

Prior works on ISAC transmit signal design exploiting distribution information assumed \emph{known} and \emph{deterministic} user locations. However, in practice, user locations can also be \emph{unknown} and \emph{random}, yet their statistical distribution can be exploited. This challenging scenario gives rise to a series of interesting problems. Firstly, with given spatial and power resources at the base station (BS) transmitter, how to design transmit beamforming to cater to different target locations and different user locations with distinct probabilities? Given the multiple possible target/user locations, is dynamic beamforming over multiple time slots  necessary? Secondly, will the ISAC performance improve as the location distributions of the user and the target become more similar or more different? Thirdly, following the previous question, in a general multi-cell system, how should each BS select its associated target and user to maximize the overall network-level ISAC performance? To the best of our knowledge, all of the above questions remain unaddressed, which motivates our study in this paper.\looseness=-1

This paper considers a MIMO ISAC system where one multi-antenna BS serves one multi-antenna communication user and senses the unknown and random location information of a point target via its reflected echo signals received at the BS. The locations (and consequently channels) for both the user and the target are unknown, while their probability distributions are known \emph{a priori}. We first establish the statistical communication and sensing performance metrics by deriving the expected communication rate and PCRB. Then, we study the transmit beamforming optimization problem to minimize the PCRB subject to an expected communication rate constraint. We derive the optimal solution and analytically prove that the rank of the optimal transmit covariance matrix is upper bounded by the summation of ranks of the MIMO communication channels corresponding to all possible user locations. We also prove that static beamforming is sufficient for achieving the optimal ISAC performance. Finally, numerical results show that the ISAC performance improves as the target/location distributions become more similar, which provides useful insights into the BS-user/-target association designs in practice.\looseness=-1

\vspace{-1mm}
\section{System Model}\label{sec_sys}
We consider a MIMO ISAC system which consists of a multi-antenna BS equipped with $N_{\mathrm{T}}\geq 1$ transmit antennas and $N_{\mathrm{R}}\geq 1$ co-located receive antennas, a multi-antenna user equipped with $N_{\mathrm{U}}\geq 1$ antennas, and a point target whose \emph{unknown} and \emph{random} location information needs to be sensed. The BS sends dual-functional signals over $L\geq 1$ symbol intervals to communicate with the user and estimate the location information of the target via the echo signals reflected by the target and received back at the BS receive antennas. Specifically, we aim to sense the target's angular location $\theta$ with respect to the BS, as illustrated in Fig. \ref{system model}. The PDF of $\theta$ denoted by $p_{\Theta}(\theta)$ is assumed to be known \emph{a priori} based on the target movement pattern or historic data \cite{xu2024mimo,xu2024isac,yao2025optimal,hou2023optimal,attiah2023active,liu2024ris,ghaddar2025active}, which will be exploited in the sensing of $\theta$.
We focus on a challenging scenario where the user's angular location $\theta_{\mathrm{U}}$ with respect to the BS and the corresponding BS-user channel are also \emph{unknown} and \emph{random}, while the distribution information is known \emph{a priori}. We assume there are $K\geq 1$ possible angular locations of the user, each denoted by $\theta_{\mathrm{U},k}$ with probability mass $p_k$, where $\sum_{k=1}^K p_k=1$. \looseness=-1

\subsection{Communication Model and Performance}
Let $\bm{H}(\theta_{\mathrm{U},k})\in \mathbb{C}^{N_{\mathrm{U}}\times N_{\mathrm{T}}}$ denote the channel matrix from the BS to the user at the angular location $\theta_{\mathrm{U},k}$, which is assumed to be perfectly known at the BS.\footnote{We assume there is a one-to-one correspondence between each possible angular location and the BS-user channel.}  Note that this is feasible when the BS-user channels follow the line-of-sight (LoS) model or under static multi-path models where $\bm{H}(\theta_{\mathrm{U},k})$'s can be analytically derived, experimentally measured, or obtained from prior communication instances.
Let $\bm{x}_l\in\mathbb{C}^{N_{\mathrm{T}}\times 1}$ denote the baseband equivalent dual-functional transmit signal vector in the $l$-th symbol interval, $l=1,...,L$. The transmit covariance matrix is thus given by $\bm{W}=\mathbb{E}[\bm{x}_l\bm{x}_l^H]$. Let $P$ denote the transmit power budget, which yields $\mathrm{tr}(\bm{W})\leq P$. \looseness=-1

At the communication user receiver, the received signal at angular location $\theta_{\mathrm{U},k}$ in each $l$-th symbol interval is given by $
\bm{y}_l^{\mathrm{C}}(\theta_{\mathrm{U},k})=\bm{H}(\theta_{\mathrm{U},k})\bm{x}_l+\bm{n}_l^{\mathrm{C}},\ l=1,...,L$,
where $\bm{n}_l^{\mathrm{C}}\sim\mathcal{CN}(\bm{0},\sigma_{\mathrm{C}}^2\bm{I}_{N_{\mathrm{U}}})$ denotes the circularly symmetric complex Gaussian (CSCG) noise in the $l$-th symbol interval, with $\sigma_{\mathrm{C}}^2$ denoting the average noise power. The achievable rate for the communication user at $\theta_{\mathrm{U},k}$ is then given by
    \begin{equation}\label{rate_narrow}
		R_k=\log_2\bigg|\bm{I}_{N_{\mathrm{U}}}+\frac{\bm{H}(\theta_{\mathrm{U},k})\bm{W}\bm{H}^H(\theta_{\mathrm{U},k})}{\sigma_{\mathrm{C}}^2}\bigg|
    \end{equation}
in bits per second per Hertz (bps/Hz).
The overall \emph{expected rate} over the $K$ possible user locations is thus given by $\sum_{k=1}^K p_k\log_2\big|\bm{I}_{N_{\mathrm{U}}}+\frac{\bm{H}(\theta_{\mathrm{U},k})\bm{W}\bm{H}^H(\theta_{\mathrm{U},k})}{\sigma_{\mathrm{C}}^2}\big|$.\looseness=-1

\subsection{Sensing Model}
By sending downlink signals, the BS can also sense $\theta$ based on the received echo signals reflected by the target and the prior distribution information about $\theta$. We consider an LoS channel between the BS and the target. The equivalent MIMO channel from the BS transmitter to the BS receiver via target reflection is given by $\alpha\bm{b}(\theta)\bm{a}^H(\theta)$.
Specifically, $\alpha\triangleq \frac{\beta_0}{r^2}\psi=\alpha_{\mathrm{R}}+j\alpha_{\mathrm{I}}\in \mathbb{C}$ denotes the reflection coefficient, which contains both the round-trip channel gain $\frac{\beta_0}{r^2}$ with $\beta_0$ denoting the reference channel power at distance $1$ m and $r$ denoting the BS-target distance in m, as well as the radar cross-section (RCS) coefficient $\psi\in\mathbb{C}$. Note that $\alpha$ is generally unknown, while its distribution can be known \emph{a priori} based on target properties. Let $\bm{a}^H(\theta)\in \mathbb{C}^{1\times N_{\mathrm{T}}}$ and $\bm{b}(\theta)\in \mathbb{C}^{N_{\mathrm{R}}\times 1}$ denote the normalized steering vectors of the transmit and receive antennas, respectively, in which $\theta$ is the only unknown parameter.
The received echo signal vector in each $l$-th symbol interval is given by \looseness=-1
    \begin{align}\label{y_l_sensing}
            \bm{y}_l^{\mathrm{S}}&=\alpha\bm{b}(\theta)\bm{a}^H(\theta)\bm{x}_l+\bm{n}_l^{\mathrm{S}},   \quad l=1,...,L,
    \end{align}
where $\bm{n}_l^{\mathrm{S}}\sim\mathcal{CN}(\bm{0}, \sigma_{\mathrm{S}}^2\bm{I}_{N_{\mathrm{R}}})$ denotes the CSCG noise vector at the BS receiver, with $\sigma_{\mathrm{S}}^2$ denoting the average noise power. The collection of received signal vectors available for sensing over $L$ symbol intervals can be expressed as $
\bm{Y}^{\mathrm{S}}=[{\bm{y}_1^{\mathrm{S}}},...,{\bm{y}_L^{\mathrm{S}}} ]=\alpha\bm{b}(\theta)\bm{a}^H(\theta)[ \bm{x}_1 ,...,\bm{x}_L ]+[{\bm{n}_1^{\mathrm{S}}},...,{\bm{n}_L^{\mathrm{S}}} ]$. Since $\alpha$ is also unknown in $\bm{Y}^{\mathrm{S}}$, $\theta$ and $\alpha$ need to be jointly sensed based on $\bm{Y}^{\mathrm{S}}$ and prior distribution information of $\theta$ and $\alpha$.

Note that the received echo signals in $\bm{Y}^{\mathrm{S}}$ and hence the estimation performance of $\theta$ are highly dependent on the transmit beamforming design. 
Moreover, the transmit beamforming design also critically affects the communication performance. Particularly, for our considered challenging scenario, the transmit signal design needs to strike an optimal balance among not only the possible target locations, but also the possible user locations based on their probability distributions. In the following, we first characterize the sensing performance as an explicit function of the transmit beamforming, based on which we formulate and solve the transmit beamforming optimization problem for ISAC when both the target and the user have unknown locations. \looseness=-1

\vspace{-1mm}
\section{Characterization of Sensing Performance Exploiting Prior Information}\label{sec_PCRB}

Let $\bm{\zeta}=[\theta, \alpha_{\mathrm{R}}, \alpha_{\mathrm{I}}]^T$ denote the collection of all unknown (real) parameters, which need to be jointly sensed to obtain an accurate estimate of $\theta$. Let $p_{\alpha_{\mathrm{R}},\alpha_{\mathrm{I}}}(\alpha_{\mathrm{R}},\alpha_{\mathrm{I}})$ denote the PDF for $\alpha$ and assume $\alpha$ is independent of $\theta$. The PDF of $\bm{\zeta}$ is given by $p_Z(\bm{\zeta})=p_{\Theta}(\theta)p_{\alpha_{\mathrm{R}},\alpha_{\mathrm{I}}}(\alpha_{\mathrm{R}},\alpha_{\mathrm{I}})$.\footnote{We consider a mild condition of $\iint\alpha_{\mathrm{R}}p_{\alpha_{\mathrm{R}},\alpha_{\mathrm{I}}}(\alpha_{\mathrm{R}},\alpha_{\mathrm{I}})d\alpha_{\mathrm{R}}d\alpha_{\mathrm{I}}=\iint\alpha_{\mathrm{I}}p_{\alpha_{\mathrm{R}},\alpha_{\mathrm{I}}}(\alpha_{\mathrm{R}},\alpha_{\mathrm{I}})d\alpha_{\mathrm{R}}d\alpha_{\mathrm{I}}=0$, which holds for various random variables (RVs), including but not limited to all proper RVs such as CSCG RVs.\looseness=-1} The posterior Fisher information matrix (PFIM) for sensing $\bm{\zeta}$ is given by $\bm{F}=\bm{F}_{\mathrm{O}}+\bm{F}_{\mathrm{P}}$, where $\bm{F}_{\mathrm{O}}\in\mathbb{R}^{3\times 3}$ denotes the PFIM from the observation in $\bm{Y}^{\mathrm{S}}$, and $\bm{F}_{\mathrm{P}}\in\mathbb{R}^{3\times 3}$ denotes the PFIM from prior information in $p_Z(\bm{\zeta})$. We assume that $L$ is sufficiently large such that the sample covariance matrix $\frac{1}{L}\sum_{l=1}^L\bm{x}_l\bm{x}_l^H$ can be accurately approximated by the transmit covariance matrix $\bm{W}$ \cite{hua2024mimo}. The PFIM $\bm{F}_{\mathrm{O}}$ from observation can be partitioned as $\bm{F}_{\mathrm{O}}=\begin{bmatrix}
	F^{\theta\theta}_{\mathrm{O}} & \bm{F}^{\theta\alpha}_{\mathrm{O}}\\
	{\bm{F}_{\mathrm{O}}^{\theta\alpha}}^H & \bm{F}^{\alpha\alpha}_{\mathrm{O}}
\end{bmatrix}$ with each block in $\bm{F}_{\mathrm{O}}$ given by $F^{\theta\theta}_{\mathrm{O}}=\frac{2L\gamma}{\sigma_{\mathrm{S}}^2}\mathrm{tr}(\bm{A}_1\bm{W})$, $\bm{F}^{\theta\alpha}_{\mathrm{O}}=\bm{0}$, $\bm{F}^{\alpha\alpha}_{\mathrm{O}}=\frac{2L}{\sigma_{\mathrm{S}}^2}\mathrm{tr}(\bm{A}_2\bm{W})\bm{I}_2$, 
where $\gamma\triangleq\iint(\alpha_{\mathrm{R}}^2+\alpha_{\mathrm{I}}^2)p_{\alpha_{\mathrm{R}},\alpha_{\mathrm{I}}}(\alpha_{\mathrm{R}},\alpha_{\mathrm{I}})d\alpha_{\mathrm{R}}d\alpha_{\mathrm{I}}$,          
		$\bm{A}_1=\int\Dot{\bm{M}}^H(\theta)\Dot{\bm{M}}(\theta)p_{\Theta}(\theta)d\theta$, $\bm{A}_2=\int\bm{M}^H(\theta)\bm{M}(\theta)p_{\Theta}(\theta)d\theta$,
	$\bm{M}(\theta)\triangleq\bm{b}(\theta)\bm{a}^H(\theta)$,
	and $\Dot{\bm{M}}(\theta)\triangleq\frac{\partial\bm{M}(\theta)}{\partial\theta}$ \cite{yao2025optimal}. Moreover, $\bm{F}_\mathrm{P}$ can be derived as $\bm{F}_{\mathrm{P}}=\Big[\mathbb{E}_\theta\Big[\Big(\frac{\partial\ln(p_{\Theta}(\theta))}{\partial\theta}\Big)^2\Big],\bm{0};\bm{0},\bm{F}^{\alpha\alpha}_{\mathrm{P}}\Big]$,
where $[\bm{F}^{\alpha\alpha}_{\mathrm{P}}]_{m,n}\!\!=\!\mathbb{E}_{\alpha_{\mathrm{R}},\alpha_{\mathrm{I}}}\Big[\frac{\partial \ln(p_{\alpha_{\mathrm{R}},\alpha_{\mathrm{I}}}(\alpha_{\mathrm{R}},\alpha_{\mathrm{I}}))}{\partial \bar{\alpha}_m }\frac{\partial \ln(p_{\alpha_{\mathrm{R}},\alpha_{\mathrm{I}}}(\alpha_{\mathrm{R}},\alpha_{\mathrm{I}}))}{\partial \bar{\alpha}_n }\!\Big]$ with $\bar{\alpha}_1=\alpha_{\mathrm{R}}, \bar{\alpha}_2=\alpha_{\mathrm{I}}$.
The overall PCRB for the MSE in estimating $\bm{\zeta}$ is given by $\bm{F}^{-1}$. The PCRB for the MSE in estimating the desired sensing parameter $\theta$ is given by \looseness=-1
\begin{align}
	\mathrm{PCRB}_{\theta}=\frac{1}{\mathbb{E}_\theta\Big[\Big(\frac{\partial\ln(p_{\Theta}(\theta))}{\partial\theta}\Big)^2\Big]+\frac{2L\gamma}{\sigma_{\mathrm{S}}^2}\mathrm{tr}(\bm{A}_1\bm{W})}.
\end{align}

Note that $\mathrm{PCRB}_{\theta}$ is determined by the transmit covariance matrix $\bm{W}$, whose optimization will be studied in Section \ref{sec_expected rate}.

\vspace{-1mm}
\section{Transmit Beamforming Optimization with Unknown Target and User Locations}\label{sec_expected rate}
\vspace{-1mm}

In this section, we establish the framework for transmit beamforming in ISAC systems with unknown target and user locations. We investigate the core trade-off between sensing and communication performance under the expected communication rate constraint, which is averaged over the user's possible locations. \looseness=-1

\vspace{-1mm}
\subsection{Problem Formulation}
\vspace{-1mm}
We aim to minimize $\mathrm{PCRB}_{\theta}$ subject to an expected communication rate constraint specified by a rate threshold of $\bar{R}>0$ bps/Hz, by optimizing the transmit beamforming design. By noting that minimizing $\mathrm{PCRB}_{\theta}$ is equivalent to maximizing $\mathrm{tr}(\bm{A}_1\bm{W})$, the optimization problem is formulated as \looseness=-1
    \begin{align}
			\!\!\!\!\!\mbox{(P1)}\, \underset{\bm{W}\succeq \bm{0}}{\max}\, & \mathrm{tr}(\bm{A}_1\bm{W}) \label{P1_obj}\\
			\mbox{s.t.}\ & \sum_{k=1}^K\!p_k\!\log_2\bigg|\bm{I}_{N_{\mathrm{U}}}\!+\!\frac{\bm{H}(\theta_{\mathrm{U},k})\bm{W}\bm{H}^H(\theta_{\mathrm{U},k})}{\sigma_{\mathrm{C}}^2}\bigg|\!\geq\! \bar{R} \label{P1_rate constraint}\\
			& \mathrm{tr}(\bm{W})\leq P \label{P1_power constraint}.
    \end{align}
Note that (P1) is a convex optimization problem, for which the optimal solution can be obtained via the interior-point method \cite{CVX} or existing software, e.g., CVX \cite{cvxtool}. To reveal the structure of the optimal solution, we leverage the Lagrange duality theory to analyze the optimal solution structure of (P1) by discussing two cases where the expected communication rate constraint in (\ref{P1_rate constraint}) is inactive or active, respectively.\looseness=-1

\addtolength{\topmargin}{0.05in}

\vspace{-1mm}
\subsection{Optimal Solution Structure of Problem (P1)}\label{sec_optimal soultion to P1}
\vspace{-1mm}

\subsubsection{Case I: Inactive expected rate constraint} 
In this case, (P1) reduces to a semi-definite program (SDP). The optimal solution to (P1) without the constraint in (\ref{P1_rate constraint}) is  $P\bm{q}_1\bm{q}_1^H$ with $\bm{q}_1\in\mathbb{C}^{N_{\mathrm{T}}\times 1}$ being the eigenvector corresponding to the largest eigenvalue of matrix $\bm{A}_1$. Denote $R_{\mathrm{S}}=\sum_{k=1}^Kp_k\log_2\big|\bm{I}_{N_{\mathrm{U}}}+\frac{P\bm{H}(\theta_{\mathrm{U},k})\bm{q}_1\bm{q}_1^H\bm{H}^H(\theta_{\mathrm{U},k})}{\sigma_{\mathrm{C}}^2}\big|$ as its corresponding rate. If $R_{\mathrm{S}}\geq \bar{R}$, the constraint in (\ref{P1_rate constraint}) is inactive and the optimal solution to (P1) is given by $\bm{W}^{\star}=P\bm{q}_1\bm{q}_1^H$. \looseness=-1

\subsubsection{Case II: Active expected rate constraint}
First, we introduce dual variables $\beta>0$ and $\mu\geq0$, which are associated with the constraints in  (\ref{P1_rate constraint}) and (\ref{P1_power constraint}), respectively. The Lagrangian of (P1) is thus given by
\begin{align}
    &\mathcal{L}(\bm{W},\beta,\mu)=\mathrm{tr}(\bm{A}_1\bm{W})+\beta\bigg(\sum_{k=1}^Kp_k\log_2\bigg|\bm{I}_{N_{\mathrm{U}}} \\
    &+\frac{\bm{H}(\theta_{\mathrm{U},k})\bm{W}\bm{H}^H(\theta_{\mathrm{U},k})}{\sigma_{\mathrm{C}}^2}\bigg|\!-\!\bar{R}\bigg)\!-\!\mu\big(\mathrm{tr}(\bm{W})-P\big),\ \bm{W}\succeq\bm{0}.\nonumber
\end{align}
The Lagrange dual function of (P1) is defined as $g(\beta,\mu)\triangleq \mathop{\mathrm{max}}\limits_{\bm{W} \succeq \bm{0}} \; \mathcal{L}(\bm{W},\beta,\mu)$. Since (P1) is convex with strong duality, we can solve it by solving its dual problem: $\underset{\beta>0,\mu\geq0}{\min}\, g(\beta,\mu)$.
Denote the optimal solution as $(\beta^{\star},\mu^{\star})$, then the matrix $\bm{W}^{\star}$ that maximizes $\mathcal{L}(\bm{W},\beta^{\star},\mu^{\star})$ is the optimal solution to (P1), which means that we can obtain $\bm{W}^{\star}$ via the following problem: \looseness=-1
\begin{align}
    \!\mbox{(P1-I)} \underset{\bm{W}\succeq\bm{0}}{\max}\!\sum_{k=1}^K\!p_k\!\log_2\!\bigg|\!\bm{I}_{N_{\mathrm{U}}}\!\!+\!\frac{\bm{H}(\theta_{\mathrm{U},k})\!\bm{W}\!\bm{H}^H\!(\theta_{\mathrm{U},k})}{\sigma_{\mathrm{C}}^2}\bigg|\!\!-\!\mathrm{tr}(\bm{Q}\bm{W}), \label{P2I_obj}
\end{align}
where $\bm{Q}=\frac{\mu^{\star}}{\beta^{\star}}\bm{I}_{N_{\mathrm{T}}}-\frac{1}{\beta^{\star}}\bm{A}_1$. For (P1-I), we have the following lemma. \looseness=-1
\begin{lemma}\label{lemma1}
    To guarantee a bounded optimal value of Problem (P1-I), $\mu^{\star}$ should satisfy $\mu^{\star}>\lambda_1$ with $\lambda_1$ denoting the largest eigenvalue of the matrix $\bm{A}_1$.
\end{lemma}

\begin{IEEEproof}
	Suppose $\mu^{\star}\leq\lambda_1$. Let $x_W\bm{q}_1\bm{q}_1^H$ with $x_W$ being any positive constant and $\|\bm{q}_1\|\neq0$ denote a feasible solution to (P1-I). Substituting $\bm{W}$ into (P1-I) yields $\sum_{k=1}^Kp_k\log_2\big(1\!+\!\frac{x_W\|\bm{H}(\theta_{\mathrm{U},k})\bm{q}_1\|^2}{\sigma_{\mathrm{C}}^2}\big)+\frac{x_W}{\beta^{\star}}(\lambda_1-\mu^{\star})$. When $x_W\rightarrow\infty$, we have $\sum_{k=1}^Kp_k\log_2\big(1+\frac{x_W\|\bm{H}(\theta_{\mathrm{U},k})\bm{q}_1\|^2}{\sigma_{\mathrm{C}}^2}\big)+\frac{x_W}{\beta^{\star}}(\lambda_1-\mu^{\star})\rightarrow\infty$ and (P1-I) becomes unbounded. Therefore, the assumption of $\mu^{\star}\leq\lambda_1$ is not true, and $\mu^{\star}>\lambda_1$ should be satisfied.
\end{IEEEproof}

Based on Lemma \ref{lemma1}, $\bm{Q}\succ\bm{0}$ holds and $\bm{Q}^{-1}$ exists. Define $\hat{\bm{W}}=\bm{Q}^{\frac{1}{2}}\bm{W}\bm{Q}^{\frac{1}{2}}\succeq\bm{0}$. (P1-I) can be rewritten as
\begin{align}
    \!\mbox{(P1-I')}\underset{\hat{\bm{W}}\succeq\bm{0}}{\max}\!&\sum_{k=1}^K\!p_k\!\log_2\!\bigg|\bm{I}_{N_{\mathrm{U}}}\!\!+\!\frac{\bm{H}\!(\theta_{\mathrm{U},k})\bm{Q}^{-\frac{1}{2}}\hat{\bm{W}}\bm{Q}^{-\frac{1}{2}}\bm{H}^H\!(\theta_{\mathrm{U},k})}{\sigma_{\mathrm{C}}^2}\bigg|\nonumber\\
    &-\mathrm{tr}(\hat{\bm{W}}). \label{P1Ieq_obj}
\end{align}
Let $f(\hat{\bm{W}})=\sum_{k=1}^Kp_k\log_2\big|\bm{I}_{N_{\mathrm{U}}}+\frac{1}{\sigma_{\mathrm{C}}^2}\bm{B}_k\hat{\bm{W}}\bm{B}_k^H\big|-\mathrm{tr}(\hat{\bm{W}})$ denote the objective function of (P1-I'), where  $\bm{B}_k=\bm{H}(\theta_{\mathrm{U},k})\bm{Q}^{-\frac{1}{2}}$. The gradient of $f(\hat{\bm{W}})$ is given by \looseness=-1
\begin{equation}\label{gradient_fW}
    \!\!\!\!\nabla f(\hat{\bm{W}})\!=\!\!\sum_{k=1}^K\!\frac{p_k}{\ln2\cdot\sigma_{\mathrm{C}}^2}\bm{B}_k^H\bigg(\bm{I}_{N_{\mathrm{U}}}\!\!+\frac{\bm{B}_k\hat{\bm{W}}\bm{B}_k^H}{\sigma_{\mathrm{C}}^2}\bigg)^{-1}\!\!\!\!\!\bm{B}_k\!-\!\bm{I}_{N_{\mathrm{T}}}.
\end{equation}
We introduce a dual variable $\bm{Z}\succeq\bm{0}$ associated with the constraint $\hat{\bm{W}}\succeq\bm{0}$. The Lagrangian of (P1-I') is thus given by $\mathcal{L}(\hat{\bm{W}},\bm{Z})=f(\hat{\bm{W}})+\mathrm{tr}(\bm{Z}\bm{W})$.
The Karush-Kuhn-Tucker (KKT) optimality conditions \cite{CVX} include the feasibility constraints and
\begin{align}
    \mathrm{tr}(\bm{Z}^{\star}\hat{\bm{W}}^{\star})&=0 \label{KKT_slack_condition}\\
    \frac{\partial \mathcal{L}(\hat{\bm{W}}^{\star},\bm{Z}^{\star})}{\partial\hat{\bm{W}}^{\star}}=\nabla f(\hat{\bm{W}}^{\star})+\bm{Z}^{\star}&=\bm{0}. \label{KKT_partial_condition}
\end{align}
Based on (\ref{KKT_partial_condition}), we have $\bm{Z}^{\star}=\bm{I}_{N_{\mathrm{T}}}-\sum_{k=1}^K\frac{p_k}{\ln2\cdot\sigma_{\mathrm{C}}^2}\bm{B}_k^H\bm{M}_k\bm{B}_k$,
where $\bm{M}_k=\Big(\bm{I}_{N_{\mathrm{U}}}+\frac{\bm{B}_k\hat{\bm{W}}^{\star}\bm{B}_k^H}{\sigma_{\mathrm{C}}^2}\Big)^{-1}\succ\bm{0}$. Then, we have the following proposition. \looseness=-1
\begin{proposition}\label{Proposition1}
    The optimal solution to (P1-I') $\hat{\bm{W}}^{\star}$ satisfies $\hat{\bm{W}}^{\star}=\bm{D}\hat{\bm{W}}^{\star}$, where $\bm{D}=\sum_{k=1}^K\frac{p_k}{\ln2}\frac{1}{\sigma_{\mathrm{C}}^2}\bm{B}_k^H\bm{M}_k\bm{B}_k$. Moreover, $\mathrm{range}(\hat{\bm{W}}^{\star})\subseteq\mathrm{range}(\bm{D})$.
\end{proposition}
\begin{IEEEproof}
    Based on (\ref{KKT_slack_condition}), we have $\bm{Z}^{\star}\hat{\bm{W}}^{\star}=\bm{0}$. Since $\bm{Z}^{\star}=\bm{I}_{N_{\mathrm{T}}}-\sum_{k=1}^K\frac{p_k}{\ln2\cdot\sigma_{\mathrm{C}}^2}\bm{B}_k^H\bm{M}_k\bm{B}_k$, we have $\hat{\bm{W}}^{\star}=\bm{D}\hat{\bm{W}}^{\star}$.  For any vector $\bm{s}\in\mathbb{C}^{N_{\mathrm{T}}\times 1}$, we have $\hat{\bm{W}}^{\star}\bm{s}=\bm{D}(\hat{\bm{W}}^{\star}\bm{s})\in\mathrm{range}(\bm{D})$. Therefore, we have $\mathrm{range}(\hat{\bm{W}}^{\star})\subseteq\mathrm{range}(\bm{D})$.
Proposition \ref{Proposition1} is thus proved.  
\end{IEEEproof}
Based on Proposition \ref{Proposition1}, we have the following proposition:
\begin{proposition}\label{Proposition2}
    When $\beta>0$, the optimal solution to (P1) satisfies $\mathrm{rank}(\bm{W}^{\star})\leq \sum_{k=1}^K\mathrm{rank}(\bm{H}(\theta_{\mathrm{U},k}))$.
\end{proposition}
\begin{IEEEproof}
    Since $\bm{D}=\sum_{k=1}^K\frac{p_k}{\ln2}\frac{1}{\sigma_{\mathrm{C}}^2}\bm{B}_k^H\bm{M}_k\bm{B}_k$, $\mathrm{rank}(\bm{D})\leq\sum_{k=1}^K\mathrm{rank}(\bm{B}_k^H\bm{M}_k\bm{B}_k)\leq\sum_{k=1}^K\mathrm{rank}(\bm{B}_k)=\sum_{k=1}^K\mathrm{rank}(\bm{H}(\theta_{\mathrm{U},k})\bm{Q}^{-\frac{1}{2}})=\sum_{k=1}^K\mathrm{rank}(\bm{H}(\theta_{\mathrm{U},k}))$. According to Proposition \ref{Proposition1}, we have $\mathrm{rank}(\hat{\bm{W}}^{\star})\leq \sum_{k=1}^K\mathrm{rank}(\bm{H}(\theta_{\mathrm{U},k}))$.
For any feasible solution $\hat{\bm{W}}$ to (P1-I'), $\bm{W}=\bm{Q}^{-\frac{1}{2}}\hat{\bm{W}}\bm{Q}^{-\frac{1}{2}}$ is a feasible solution to (P1-I) with the same objective value. Therefore, we have $\mathrm{rank}(\bm{W}^{\star})=\mathrm{rank}(\bm{Q}^{-\frac{1}{2}}\hat{\bm{W}}^{\star}\bm{Q}^{-\frac{1}{2}})=\mathrm{rank}(\hat{\bm{W}}^{\star})\leq \sum_{k=1}^K\mathrm{rank}(\bm{H}(\theta_{\mathrm{U},k}))$.
The proof of Proposition \ref{Proposition2} is thus completed. \looseness=-1
\end{IEEEproof}

These results indicate that the rank of the optimal transmit covariance matrix is bounded by the sum of the ranks of the MIMO communication channels over all possible user locations, even though the transmit signals need to simultaneously cater to a multi-antenna communication user and a sensing target with an infinite number of possible locations. \looseness=-1

\section{Impact of Temporal Flexibility} \label{sec_multi_slot}

The designs in Section \ref{sec_expected rate} assumed a static transmit strategy over the entire ISAC duration of $L$ symbol intervals.
A natural and important question thus arises: can dynamically allocating resources across time slots offer additional performance benefits? This temporal flexibility introduces new degrees of freedom for system design, which may be needed for our considered challenging scenario with user location uncertainty, as studied in the following.\looseness=-1

\vspace{-1mm}
\subsection{System Model with Temporal Flexibility}
\vspace{-1mm}

With a slight abuse of notation, we consider $L$ consecutive transmission blocks indexed by $l=1,..., L$, where each block duration $T_s$ matches the symbol interval duration in Section \ref{sec_sys}. To enhance the design flexibility, 
we introduce $M$ equal-sized time slots indexed by $m=1,..., M$ within each transmission block. The baseband equivalent dual-functional transmit signal vector in the $m$-th time slot of the $l$-th transmission block is denoted by $\bm{x}_{l,m}\in\mathbb{C}^{N_{\mathrm{T}}\times 1}$.\looseness=-1

The transmit signal design now involves $M$ transmit covariance matrices, each corresponding to a specific time slot within the transmission block. For each time slot $m$, the transmit covariance matrix is defined as $\bm{W}_m=\mathbb{E}[\bm{x}_{l,m}\bm{x}_{l,m}^H],\ m=1,..., M$, which remains invariant across all transmission blocks $l$ due to the quasi-static nature of the channel and target. 

At the communication user receiver, the received signal at the location $\theta_{\mathrm{U},k}$ in the $m$-th time slot of each $l$-th transmission block is given by $\bm{y}_{l,m}^{\mathrm{C}}(\theta_{\mathrm{U},k})\!=\!\bm{H}(\theta_{\mathrm{U},k})\bm{x}_{l,m}+\bm{n}_{l,m}^{\mathrm{C}},\ l=1,..., L,\ m=1,... ,M$,
where $\bm{n}_{l,m}^{\mathrm{C}}\sim\mathcal{CN}(\bm{0},\sigma_{\mathrm{C}}^2\bm{I}_{N_{\mathrm{U}}})$ denotes the CSCG noise in the $m$-th time slot of the $l$-th transmission block. Since the channel remains constant across all time slots and transmission blocks, and the noise is independent across time slots, the user can jointly decode the information from all $M$ time slots within each transmission block. Since each time slot occupies only $\frac{1}{M}$-th of the transmission block duration, the achievable rate for the communication user at location $\theta_{\mathrm{U},k}$ within each transmission block is given by $R_{k}^{\mathrm{MTS}}=\frac{1}{M}\sum_{m=1}^M\log_2\big|\bm{I}_{N_{\mathrm{U}}}+\frac{\bm{H}(\theta_{\mathrm{U},k})\bm{W}_m\bm{H}^H(\theta_{\mathrm{U},k})}{\sigma_{\mathrm{C}}^2}\big|$
in bps/Hz. The \emph{expected rate} is given by 
$R_{\mathrm{E}}^{\mathrm{MTS}}=\frac{1}{M}\sum_{k=1}^K \sum_{m=1}^Mp_k\log_2\big|\bm{I}_{N_{\mathrm{U}}}+\frac{\bm{H}(\theta_{\mathrm{U},k})\bm{W}_m\bm{H}^H(\theta_{\mathrm{U},k})}{\sigma_{\mathrm{C}}^2}\big|$.

By sending downlink signals, the BS can also sense $\theta$ based on the echo signals reflected by the target and the prior distribution information about $\theta$. The received echo signal vector in the $m$-th time slot of each $l$-th transmission block is given by \looseness=-1
    \begin{align}\label{y_l_sensing_MTS}
            \bm{y}_{l,m}^{\mathrm{S}}\!\!=\!\alpha\bm{b}(\theta)\bm{a}^H(\theta)\bm{x}_{l,m}\!+\!\bm{n}_{l,m}^{\mathrm{S}},   l\!=\!1,\!...,\!L, m\!=\!1,\!...,\!M,
    \end{align}
where $\bm{n}_{l,m}^{\mathrm{S}}\sim\mathcal{CN}(\bm{0}, \sigma_{\mathrm{S}}^2\bm{I}_{N_{\mathrm{R}}})$ denotes the CSCG noise vector at the BS receiver. The collection of received signal vectors available for sensing over $L$ transmission blocks and $M$ time slots is given by $\bm{Y}^{\mathrm{S}}_{\mathrm{MTS}}\!\!=\![{\bm{y}_{1,1}^{\mathrm{S}}},...,{\bm{y}_{1,M}^{\mathrm{S}}}\big|\!...\big|{\bm{y}_{L,1}^{\mathrm{S}}},...,{\bm{y}_{L,M}^{\mathrm{S}}} ]\!\!=\!\alpha\bm{b}(\theta)\bm{a}^H(\theta)\bm{X}\!\!+\!\bm{N}$,
with $\bm{X}=[\bm{x}_{1,1},...,\bm{x}_{L,M}]$ and $\bm{N}$ denoting the corresponding transmit signal and noise matrices. Since $\alpha$ is also unknown in $\bm{Y}^{\mathrm{S}}_{\mathrm{MTS}}$, $\theta$ and $\alpha$ need to be jointly sensed based on $\bm{Y}^{\mathrm{S}}_{\mathrm{MTS}}$ and prior distribution information of $\theta$ and $\alpha$.

We aim to characterize the sensing performance with multiple time slots. We assume that $L$ is sufficiently large such that the sample covariance matrix $\frac{1}{L}\sum_{l=1}^L\bm{x}_{l,m}\bm{x}_{l,m}^H$ can be accurately approximated by the transmit covariance matrix $\bm{W}_m$ \cite{hua2024mimo}. The PCRB for the MSE in estimating the desired parameter $\theta$ with multiple time slots is thus given by \looseness=-1
    \begin{align}\label{PCRB_theta_MTS}
    	\!\!\!
		\mathrm{PCRB}_{\theta}\!\!=\!\frac{1}{\mathbb{E}_\theta\Big[\!\Big(\!\frac{\partial\ln(p_{\Theta}(\theta))}{\partial\theta}\!\Big)^{\!\!2}\Big]\!\!+\!\!\frac{2L\gamma}{\sigma_{\mathrm{S}}^2M}\!\sum_{m=1}^M\!\mathrm{tr}(\bm{A}_1\bm{W}_m)}\!.
	\end{align}

\vspace{-1mm}
\subsection{Beamforming Optimization with Temporal Flexibility for MIMO ISAC}
\vspace{-1mm}

With the additional temporal flexibility, we formulate the PCRB minimization problem under expected rate constraint as follows:\looseness=-1
  \begin{align}
	\mbox{(P2)}\underset{\{\bm{W}_m\succeq \bm{0}\}_{m=1}^M}{\max} & \sum_{m=1}^M\mathrm{tr}(\bm{A}_1\bm{W}_m) \label{P2_obj}\\
	\mbox{s.t.}\ \ \quad & R_{\mathrm{E}}^{\mathrm{MTS}}
	\geq \bar{R} \label{P2_rate constraint}\\            
	& \frac{1}{M}\sum_{m=1}^M\mathrm{tr}(\bm{W}_m)\leq P \label{P2_power constraint}.
\end{align}
Note that (P2) is a convex optimization problem. We then analyze the properties of the optimal transmit covariance matrices on multiple time slots to draw useful insights. \looseness=-1
\begin{proposition}\label{Proposition4}
    For Problem (P2), a static beamforming strategy is in fact optimal. Specifically, the optimal solution is $\bm{W}_m^\star=\bm{W}^{\star},\ \forall m$,  where $\bm{W}^{\star}$ denotes the optimal solution to (P1). \looseness=-1
\end{proposition}
\begin{IEEEproof}
	Let $\{\bm{W}_m^{\star}\}_{m=1}^M$ be an arbitrary set of optimal solutions to (P2). By definition, this solution set is feasible, i.e., it satisfies the constraints in (\ref{P2_rate constraint}) and (\ref{P2_power constraint}). We construct a new static solution denoted by $\{\bm{W}_m'\}_{m=1}^M$ where $\bm{W}_m'=\bar{\bm{W}}=\frac{1}{M}\sum_{m=1}^M\bm{W}_m^{\star},\ \forall m$. We prove Proposition \ref{Proposition4} by showing that this new solution is feasible and optimal for (P2). \looseness=-1

First, note that the optimal value of (P2) can be achieved since  $\sum_{m=1}^M\mathrm{tr}(\bm{A}_1\bm{W}_m')=M\mathrm{tr}(\bm{A}_1\bar{\bm{W}})=M\mathrm{tr}(\bm{A}_1(\frac{1}{M}\sum_{m=1}^M\bm{W}_m^\star))=\sum_{m=1}^M\mathrm{tr}(\bm{A}_1\bm{W}_m^\star)$. Then, we have $\frac{1}{M}\sum_{m=1}^M\mathrm{tr}(\bm{W}_m')=\mathrm{tr}(\bar{\bm{W}})=\mathrm{tr}(\frac{1}{M}\sum_{m=1}^M\bm{W}_m^{\star})=\frac{1}{M}\sum_{m=1}^M\mathrm{tr}(\bm{W}_m^{\star})\leq P$, i.e., the constraint in (\ref{P2_power constraint}) is satisfied. 
Moreover, denote $f_k(\bm{W})=\log_2\big|\bm{I}_{N_{\mathrm{U}}}+\frac{\bm{H}(\theta_{\mathrm{U},k})\bm{W}\bm{H}^H(\theta_{\mathrm{U},k})}{\sigma_{\mathrm{C}}^2}\big|$, which is a concave function with respect to $\bm{W}$. The expected rate for the new static solution is given by $\sum_{k=1}^Kp_k(\frac{1}{M}\sum_{m=1}^Mf_k(\bar{\bm{W}}))=\sum_{k=1}^Kp_kf_k(\bar{\bm{W}})$. By Jensen's inequality, for each $k$, we have $f_k(\bar{\bm{W}})=f_k(\frac{1}{M}\sum_{m=1}^M\bm{W}_m^{\star})\geq\frac{1}{M}\sum_{m=1}^Mf_k(\bm{W}_m^{\star})$. Since $p_k\geq0$, we have $\sum_{k=1}^Kp_kf_k(\bar{\bm{W}})\geq\sum_{k=1}^Kp_k(\frac{1}{M}\sum_{m=1}^Mf_k(\bm{W}_m^{\star}))\geq \bar{R}$, 
i.e., the constraint in (\ref{P2_rate constraint}) is satisfied. Finally, since $\bm{W}_m'=\frac{1}{M}\sum_{m=1}^M\bm{W}_m^{\star},\ \forall m$, $\bm{W}_m'\succeq\bm{0}$ holds. Therefore, the new static solution is also an optimal solution to (P2), which completes the proof of Proposition \ref{Proposition4}.\looseness=-1
\end{IEEEproof}

The results of Proposition \ref{Proposition4} shows that extra temporal flexibility is not needed for the considered scenario to achieve an optimal average-sense balance between the sensing performance and statistical communication performance. Therefore, optimizing a static transmit covariance matrix is already sufficient, for which the complexity is much smaller than that in the case exploiting temporal flexibility.\looseness=-1

\vspace{-1mm}
\section{Numerical Results}\label{sec_numerical results}

In this section, we provide numerical results to evaluate our proposed beamforming designs. Unless specified otherwise, we set $N_{\mathrm{T}}=10$, $N_{\mathrm{R}}=12$, $N_{\mathrm{U}}=8$, $K=100$, $L=25$, $P=30$ dBm, $\sigma_{\mathrm{C}}^2=\sigma_{\mathrm{S}}^2=-90$ dBm, $\bar{R}=12$ bps/Hz, and $\alpha\sim\mathcal{CN}(0,2\times 10^{-14})$. We consider a uniform linear
array (ULA) layout for all antenna arrays with half-wavelength spacing, which yields $\bm{a}(\theta)=[e^{\frac{-j\pi (N_{\mathrm{T}}-1)\sin\theta}{2}},...,e^{\frac{j\pi (N_{\mathrm{T}}-1)\sin\theta}{2}}]^T$ and $\bm{b}(\theta)=[e^{\frac{-j\pi (N_{\mathrm{R}}-1)\sin\theta}{2}},...,e^{\frac{j\pi (N_{\mathrm{R}}-1)\sin\theta}{2}}]^T$. We assume that the PDF of $\theta$ follows a Gaussian PDF, which is given by $p_{\Theta}(\theta)=\frac{1}{\sqrt{2\pi}\sigma_{\mathrm{T}}}e^{-\frac{(\theta-\hat{\theta}_{\mathrm{T}})^2}{2\sigma_{\mathrm{T}}^2}}$. Specifically, $\hat{\theta}_{\mathrm{T}}\in[-\frac{\pi}{2}, \frac{\pi}{2})$ and $\sigma_{\mathrm{T}}^2$ are the mean and variance of the Gaussian PDF, respectively. \looseness=-1

For ease of comparison with the unknown target locations, we assume the probability mass function (PMF) of $\theta_{\mathrm{U}}$ is the discretized version of a 
Gaussian PDF, which is given by $p_{\Theta_{\mathrm{U}}}(\theta_{\mathrm{U}})=\frac{1}{\sqrt{2\pi}\sigma_{\mathrm{U}}}e^{-\frac{(\theta_{\mathrm{U}}-\hat{\theta}_{\mathrm{U}})^2}{2\sigma_{\mathrm{U}}^2}}$. Specifically, the user's angular domain $\Theta_{\mathrm{U}}$ is partitioned into $K$ disjoint intervals $\{\mathcal{I}_k\}_{k=1}^K$ such that $\bigcup_{k=1}^K\mathcal{I}_k=\Theta_{\mathrm{U}}$, $\mathcal{I}_i\cap\mathcal{I}_j=\emptyset$, $\forall i\neq j$. A representative angle $\theta_{\mathrm{U},k}$ is assigned to each interval $\mathcal{I}_k$ as its center. The discrete probability mass $p_k\in[0,1]$ associated with $\theta_{\mathrm{U},k}$ is given by $p_k=\int_{\mathcal{I}_k}p_{\Theta_{\mathrm{U}}}(\theta_{\mathrm{U}})d\theta_{\mathrm{U}},\ k=1,...,K$ with $\sum_{k=1}^Kp_k=1$. The BS-user channel, at a distance of $r_{\mathrm{U}}=500$ m, is modeled by a geometric channel consisting of one LoS path and $N_{\mathrm{sc}}=8$ non-LoS (NLoS) paths with $\bm{H}(\theta_{\mathrm{U},k})=\bm{H}_{\mathrm{LoS}}(\theta_{\mathrm{U},k})+\bm{H}_{\mathrm{NLoS},k}$. The LoS component is modeled as
 $\bm{H}_{\mathrm{LoS}}(\theta_{\mathrm{U},k})=\sqrt{\beta_{\mathrm{C}}}\bm{b}_{\mathrm{U}}(\theta_{\mathrm{U},k})\bm{a}^H(\theta_{\mathrm{U},k})$, where $\beta_{\mathrm{C}}=\frac{\beta_0}{r_{\mathrm{U}}^{3.2}}$ denotes the path power gain with $\beta_0=-30$ dB, and $\bm{b}_{\mathrm{U}}(\theta_{\mathrm{U},k})$ denotes the steering vector of the user receive antenna array. The NLoS component comprises the sum of the $N_{\mathrm{sc}}$ paths with $\bm{H}_{\mathrm{NLoS},k}=\sum_{n=1}^{N_{\mathrm{sc}}}\eta_{k}^n\bm{b}_{\mathrm{U}}(\phi_{\mathrm{U},k}^n)\bm{a}^H(\phi_{\mathrm{T},k}^n)$, where the angle-of-arrivals (AoAs) $\phi_{\mathrm{U},k}^n$ and angle-of-departures (AoDs) $\phi_{\mathrm{T},k}^n$ for each path are independent and identically distributed random variables, and the complex gain is distributed as $\eta_{k}^n\sim\mathcal{CN}(0, \frac{\beta_{\mathrm{C}}}{\Lambda_{\mathrm{C}}N_{\mathrm
 		{sc}}})$ with $\Lambda_{\mathrm{C}}=0.8$ dB representing the power ratio between the LoS and NLoS components. For comparison, we consider the following benchmark schemes. \looseness=-1
\begin{itemize}
	\item \textbf{Benchmark Scheme 1:} This scheme divides the $L$ symbol intervals into two phases. In the first phase, $L_{\mathrm{CE}}<L$ symbols are used for channel estimation. An optimal training matrix $\bm{P}$, with $\bm{P}\bm{P}^H\propto\bm{I}_{N_{\mathrm{T}}}$ \cite{biguesh2006training} is transmitted to estimate the BS-user channel $\bm{H}(\theta_{\mathrm{U}})$ via least squares (LS) estimation. In the second phase, using the estimated channel, the BS designs the transmit covariance for the remaining $L_{\mathrm{ISAC}}=L-L_{\mathrm{CE}}$ symbols to perform ISAC, following the methodology in \cite{xu2024mimo} for a known communication channel. \looseness=-1
	\item \textbf{Benchmark Scheme 2:} This scheme also employs a two-phase approach similar to Benchmark Scheme 1. The key difference is that the pilot signals transmitted during the initial $L_{\mathrm{CE}}$ symbols are utilized for \emph{both} channel estimation and sensing. While these pilots are structured to be optimal for LS channel estimation, this structure is generally suboptimal for the sensing task. \looseness=-1
\end{itemize}

\begin{figure}[t]
	\centering
	\includegraphics[width=0.48\textwidth]{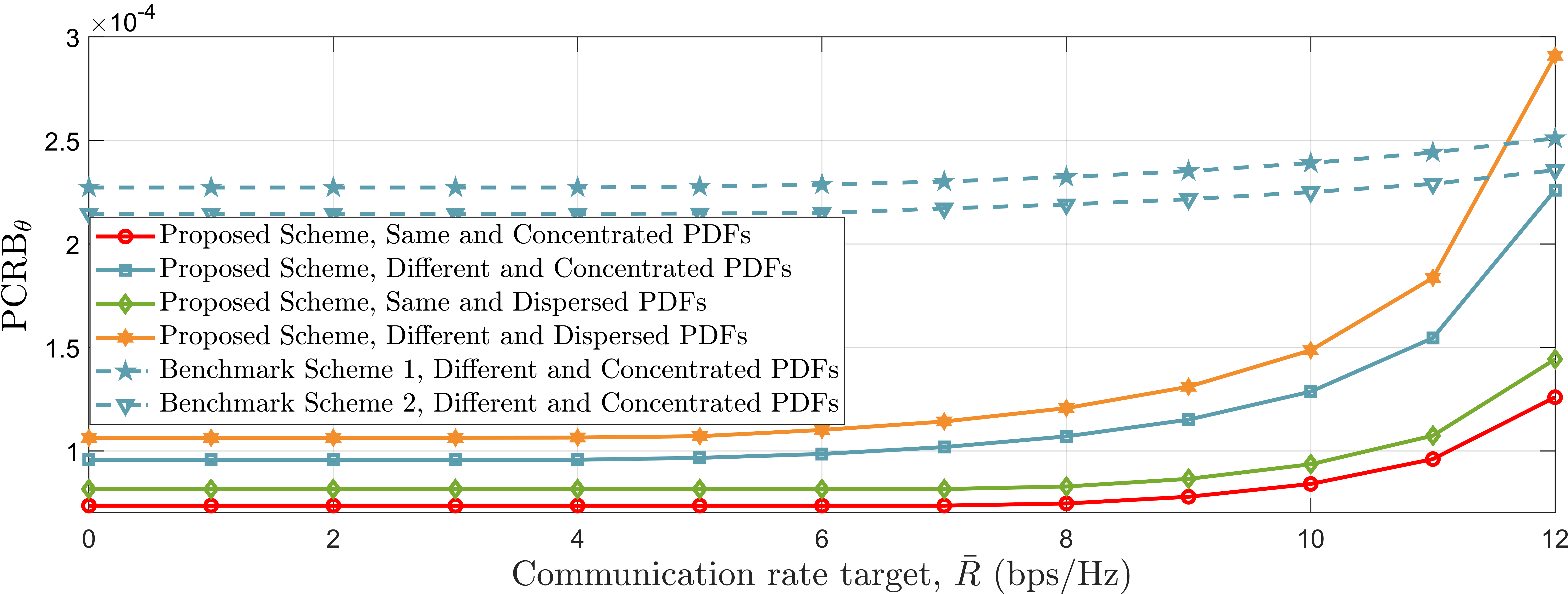}
	\vspace{-4mm}
	\caption{PCRB versus communication rate target.}
	\label{PCRB_vs_R}
\end{figure}
Fig. \ref{PCRB_vs_R} illustrates the PCRB versus the communication rate target for different target/user location distributions. For the benchmark schemes, we set $L_{\mathrm{CE}}=10$ and $L_{\mathrm{ISAC}}=15$.
We set $\hat{\theta}_{\mathrm{T}}=\hat{\theta}_{\mathrm{U}}=-0.3$ for the same PDFs and $\hat{\theta}_{\mathrm{T}}=-0.6$, $\hat{\theta}_{\mathrm{U}}=-0.3$ for different PDFs; $\sigma_{\mathrm{T}}^2=\sigma_{\mathrm{U}}^2=10^{-3}$ for concentrated PDFs and $\sigma_{\mathrm{T}}^2=\sigma_{\mathrm{U}}^2=10^{-2.5}$ for dispersed PDFs. 
It is observed that the PCRB increases as the communication rate target becomes larger for the different PDFs, due to the necessity of allocating the limited power and spatial resources at the BS transmitter between the communication and sensing functions, which leads to the non-trivial PCRB-rate trade-off. 
Furthermore, concentrated user/target PDFs yield superior PCRB performance to dispersed PDFs, as concentrated angular distributions prevent the dilution of limited power and spatial resources, enabling focused beamforming toward high-probability angles. Moreover, our proposed scheme outperforms both benchmarks. This is because our approach utilizes the full $L$ symbols for joint beamforming design, whereas the benchmarks dedicate $L_{\mathrm{CE}}$ symbols to channel estimation. Even when these pilot symbols are reused for sensing (Benchmark Scheme 2), their structure is tailored for estimation and thus suboptimal for sensing, leading to a performance loss.\looseness = -1

\begin{figure}[t]
	\centering
	\begin{minipage}{0.49\linewidth}
		\centering
		\includegraphics[width=0.99\linewidth]{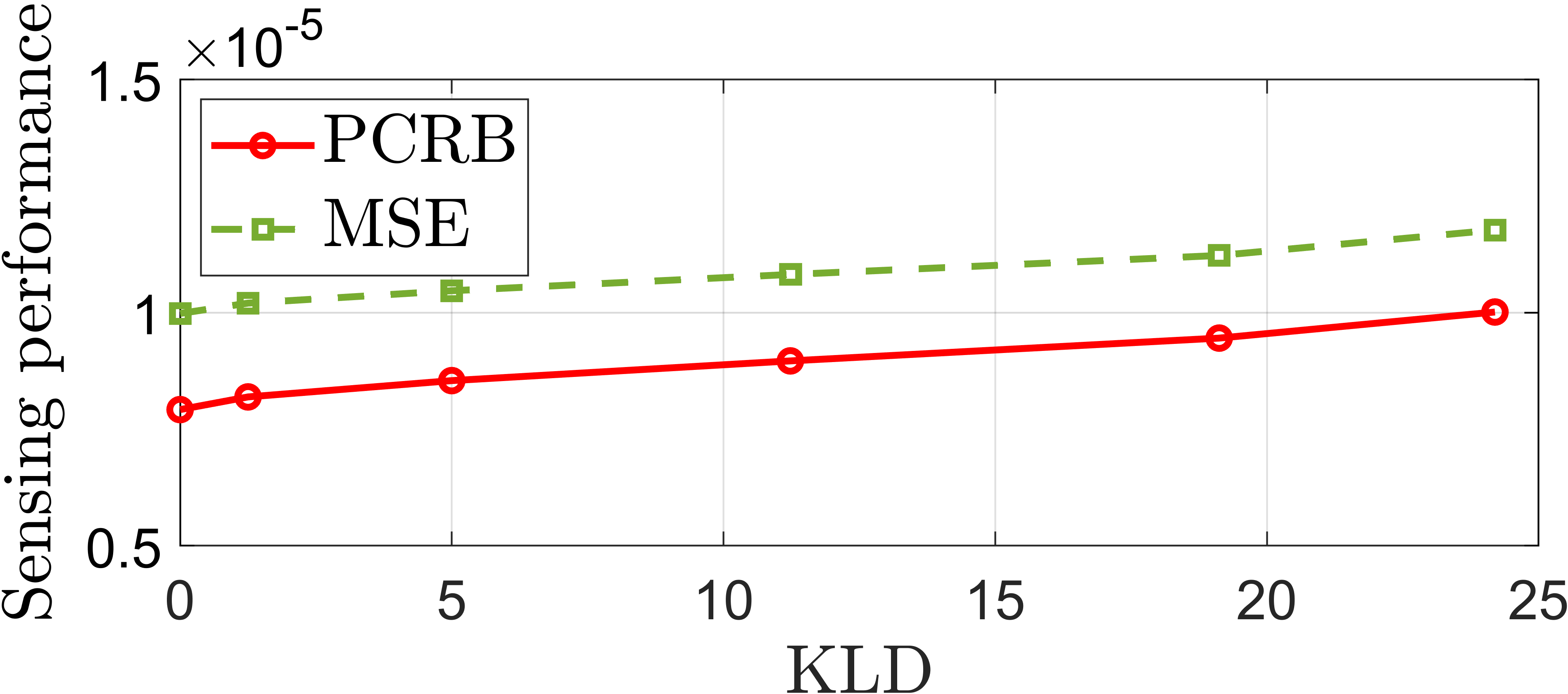}
        \vspace{-8mm}
		\caption{MSE/PCRB versus KLD for Gaussian distribution.}
		\label{MSE_Vs_KLD}
	\end{minipage}
	\begin{minipage}{0.49\linewidth}
		\centering
		\includegraphics[width=0.99\linewidth]{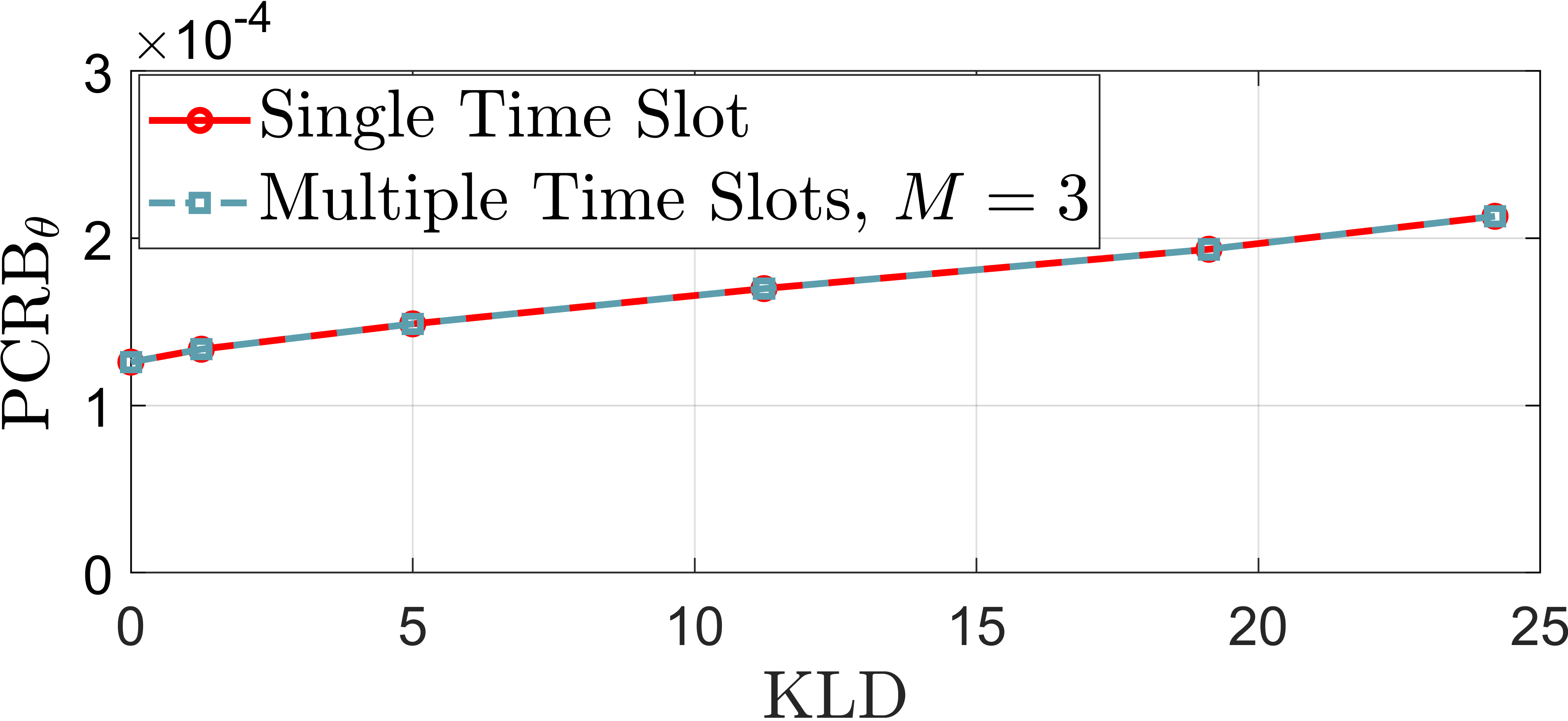}
        \vspace{-8mm}
		\caption{PCRB versus KLD with/without temporal flexibility.}
		\label{PCRB_Vs_KLD_multi_time_slots}
	\end{minipage}
\end{figure}

Then, we fix the user's PDF and evaluate the MSE and the PCRB versus the Kullback-Leibler divergence (KLD) between the user's and target's PDFs in Fig. \ref{MSE_Vs_KLD} under $P=40$ dBm. The KLD quantifies the spatial dissimilarity between the two probability distributions, which is defined as $\int_{-\infty}^{+\infty}p_{\Theta_{\mathrm{U}}}(\theta)\ln\frac{p_{\Theta_{\mathrm{U}}}(\theta)}{p_{\Theta}(\theta)}d\theta$.
It is shown that both the MSE and the PCRB increase with KLD. This demonstrates that as the high-probability regions for the user and the target become more spatially separated, the transmitter must compromise by splitting its resources, leading to a degradation in sensing performance. The proximity of the MSE to PCRB validates PCRB as a tight and reliable performance metric for our optimization framework.
Fig. \ref{PCRB_Vs_KLD_multi_time_slots} validates our theoretical finding in Section \ref{sec_multi_slot} that introducing temporal flexibility with time-varying covariance matrices offers no performance gain over a static strategy under the expected rate constraint. 

\begin{figure}[t]
	\centering
	\vspace{-4mm}
	\includegraphics[width=0.41\textwidth]{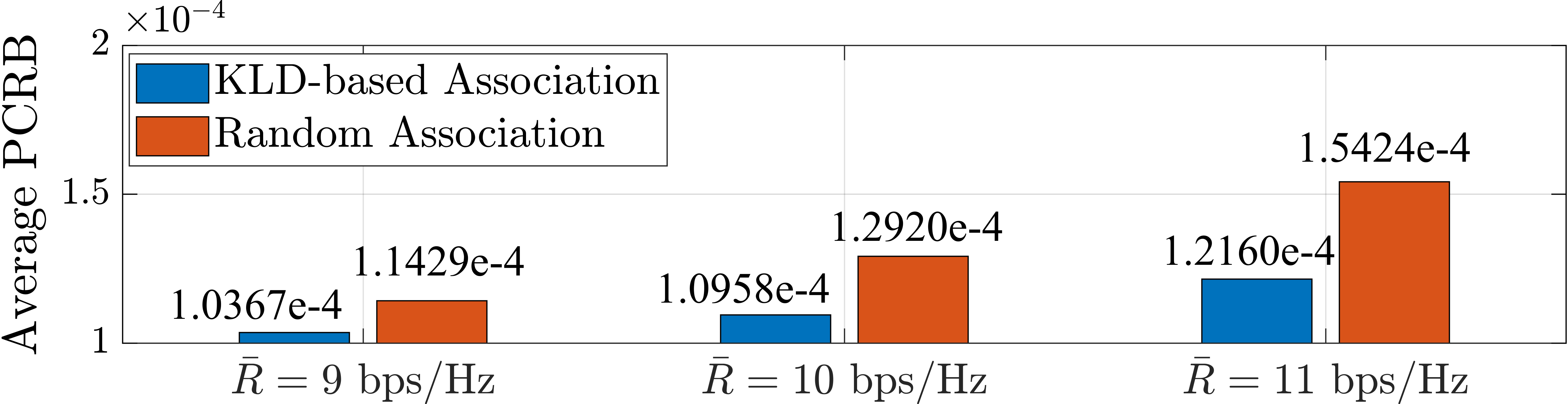}
	\vspace{-4mm}
	\caption{PCRB performance with different BS-user/-target association.}
	\label{association}
\end{figure}
Finally, we extend our analysis to a multi-cell scenario to investigate the impact of BS-user/-target association strategies. We consider a network with $N=8$ cells, each to be associated with one BS, one communication user, and one sensing target. Each user and target is characterized by a unknown location with a unique Gaussian PDF. The challenge is to decide which target and user should be associated with BS to optimize the overall network-level ISAC performance.
To address this, we propose a KLD-based association algorithm which computes a cost matrix of KLDs between user and target PDFs and then pairs them to minimize spatial dissimilarity within each ISAC group. This strategy is compared against a random association benchmark, with performance evaluated by the average PCRB across all cells.
\looseness=-1

Fig. \ref{association} shows the PCRB performance with different association strategies. It is observed that the KLD-based association strategy achieves a lower average PCRB compared to the random association. This superiority stems from the KLD-based algorithm intelligently pairs users and targets with high spatial proximity. This enables the beamforming optimization to concentrate energy on the high-probability angular regions shared by both the user and the target, resulting in a more efficient allocation of resources and improved sensing performance.\looseness=-1

\vspace{-1mm}
\section{Conclusions}

This paper developed a framework for optimal transmit beamforming design in MIMO ISAC systems with unknown and random target and user locations. By leveraging the PCRB as the performance metric, we first analyzed the non-trivial trade-offs between sensing and communication under an expected rate constraint and derived the optimal transmit beamforming solution. Then, we proved that a static, time-invariant beamforming strategy is optimal, and dynamic beamforming over multiple time slots is not needed. This insight stems from the time-averaged nature of the rate constraint. Finally, numerical results demonstrated the effectiveness of the proposed beamforming design for MIMO ISAC systems and provided useful insights on the relationship between ISAC performance and target/user location distribution similarity. \looseness=-1

\bibliographystyle{IEEEtran}
\bibliography{reference}

\end{document}